# Correlation coefficients and Robertson-Schroedinger uncertainty relations

Gordon N. Fleming
*104 Davey Lab, Pennsylvania State Univ., Univ. Park, PA 16803*
*gnf1@earthlink.net*

**Abstract:** Calling the quantity; $2\Delta A \Delta B/|<[A, B]>|$, with non-zero denominator, the *uncertainty product ratio* or UPR for the pair of observables, (A, B), it is shown that *any* non-zero correlation coefficient between two observables raises, above unity, the lower bound of the UPR for each member of an *infinite collection* of pairs of incompatible observables. Conversely, *any* UPR is subject to lower bounds above unity determined by each of an *infinite collection* of correlation coefficients. This result generalizes the well known Schroedinger strengthening of the Robertson uncertainty relations (with the former expressed in terms of the correlation coefficient rather than the anticommutator) where the UPR and the correlation coefficient both involve the same pair of observables. Two, independent, derivations of the result are presented to clarify its' origins and some examples of its' use are examined.

**1. Introduction:** The well known Robertson (1929, 1930) -Schroedinger (1931) uncertainty relation for observables, A and B, when expressed in terms of the correlation coefficient, $K(A, B)$, rather than the anti-commutator expectation value, $<\{(A - <A>), (B - <B>)\}>$, takes the form,

(1.1) $\qquad 2 \Delta A \, \Delta B \geq | <[A, B]>/(1 - K(A, B)^2)^{1/2} |$ ,

where, for any observables, X and Y, $\Delta X$ and $\Delta Y$ are the rms deviations and the correlation coefficient (CC), $K(X, Y)$, is defined by,

(1.2) $\qquad 2 \Delta X \, \Delta Y \, K(X, Y) \equiv <\{(X - <X>), (Y - <Y>)\}>$.

It will be shown here that the inequality in (1.1) continues to hold whenever the CC between A and B is replaced by a CC between A and *any observable 'effectively' compatible with* B or by a CC between B and *any observable 'effectively' compatible with* A. The term, 'effectively' compatible indicates the vanishing of only the expectation value of a commutator. In other words, if, in the right hand side of (1.1), $K(A, B)$ is replaced by $K(A, C)$ or by $K(D, B)$, where C is *any* observable satisfying $<[B, C]> = 0$ and D is *any* observable satisfying $<[A, D]> = 0$, then we still have,

(1.3a) $\qquad 2 \Delta A \, \Delta B \geq | <[A, B]>/(1 - K(A, C)^2)^{1/2} |$ ,

and

(1.3b) $\qquad 2 \Delta A \, \Delta B \geq | <[A, B]>/(1 - K(D, B)^2)^{1/2} |$ .

This means that the lower bound on the ratio,

(1.4) $\quad\quad\quad\quad\quad\quad\quad 2\,\Delta A\,\Delta B\,/\,|\langle[A, B]\rangle|$

may be raised by searching through the (effective) commutants of B or A for observables, C or D, respectively, yielding the highest squared CCs with A or B, respectively. Throughout this paper a ratio of the form (1.4), with non-zero denominator, will be called an *uncertainty product ratio* (UPR), in this case for A and B.

Conversely, a non-zero value for a CC between any two observables, even compatible ones, raises above unity the lower bound for the UPRs of *an infinite collection* of pairs of mutually incompatible observables, and not merely, as with Schroedinger, the UPR for the observables of the said CC.

A natural question is whether there exist in the (effective) commutants of B or A, observables C or D, respectively, that yield CCs that exhaust the inequality in (1.3) and achieve equality. It will be shown that the answer is; in some cases yes, but not generally. Explicit examples of both cases will be displayed in section **5**. A little reflection shows that a general affirmative answer to the question is not to be expected, since, in any case where equality is achieved, say by using the CC, $K(A, C)$, the UPR for the pair, (A, B), must then be not greater than the UPR for any pair, (A, E), where E is effectively compatible with C, nor for any pair, (D, C), where D is effectively compatible with A. Indeed, it may be precisely such minimum UPRs for which appropriate CCs *can* be found to force equality in (1.3), but that issue will not be addressed here.

Two distinct derivations of (1.3) will be presented. The first, which originally led the author to the result, is based on the *master inequality* demonstrated some time ago (Fleming, 2001). It will show (1.3) to be a special case of an *asymmetric* uncertainty relation for arbitrary sets of three observables that will be derived in section **2**, viz.,

(1.5) $\quad\quad 4\,\Delta A^2\,\Delta B^2\,\Delta C^2 \;\underline{\geq}\; \langle i[A, B]\rangle^2\,\Delta C^2 + \langle i[B, C]\rangle^2\,\Delta A^2$

$\quad\quad\quad\quad\quad\; +\,\langle\{DC, DA\}\rangle^2\,\Delta B^2 + \langle i[A, B]\rangle\langle i[B, C]\rangle\langle\{DC, DA\}\rangle,$

where, for any observable, X,

(1.6) $\quad\quad\quad\quad\quad\quad\quad\quad DX \;\underline{=}\; X - \langle X\rangle.$

In passing we will comment briefly on the relation of (1.5) to earlier derivations of uncertainty relations for more than two observables. The special case of (1.5) that yields (1.3) will be discussed in section **3**.

The second derivation, in section **4**, will be very specific to the connection between the CCs, $K(A, B)$, $K(A, C)$ and $K(D, B)$, with C and D effectively compatible with B and A, respectively. It is based on identifying the CCs as real parts of inner products of particular unit norm state vectors.

While our derivations all work with pure states, the results (1.5) and (1.3) hold for mixed states as well. This follows since both results assert a relationship among expectation values of various operators for a single state and all such expectation values for a single mixed state can, as is well known, always be expressed as corresponding expectation values of a single pure state in a larger Hilbert space.

In section **5** we discuss some simple and illuminating examples of (1.3).

**2. Derivation of (1.5) via the *master inequality*:** For pure states the *master inequality*, from which we begin, is as follows: Let $|\psi\rangle$ and $|\psi'\rangle$ denote unit norm vectors in the quantum state space and let A be *any* observable defined on both vectors. Then, if we write,

(2.1) $$|\langle\psi|\psi'\rangle| \equiv \cos\theta, \qquad 0 \leq \theta \leq \pi/2,$$

we must have

(2.2) $$\frac{|\langle A\rangle' - \langle A\rangle|}{\Delta A' + \Delta A} \leq \tan\theta,$$

where

(2.3) $$\langle A\rangle \equiv \langle\psi|A|\psi\rangle, \qquad \langle A\rangle' \equiv \langle\psi'|A|\psi'\rangle,$$
$$\Delta A^2 \equiv \langle(DA)^2\rangle, \quad \Delta A'^2 \equiv \langle(D'A)^2\rangle'.$$

From this we next derive (1.5). Consider the case in which $|\psi'\rangle$ is obtained from $|\psi\rangle$ by the transformation,

(2.4) $$|\psi'\rangle \equiv (I + i\beta\, DB + \gamma\, DC)|\psi\rangle / [1 + \beta^2\, \Delta B^2 + \gamma^2\, \Delta C^2 - \beta\gamma\langle i[B, C]\rangle]^{1/2},$$

where B and C are observables defined on $|\psi\rangle$, $\beta$ and $\gamma$ are real and the square root in the denominator is non-negative. With this definition of $|\psi'\rangle$ it follows that,

(2.5a) $$\cos\theta = 1 / [1 + \beta^2\, \Delta B^2 + \gamma^2\, \Delta C^2 - \beta\gamma\langle i[B, C]\rangle]^{1/2},$$

and

(2.5b) $$\tan\theta = [\beta^2\, \Delta B^2 + \gamma^2\, \Delta C^2 - \beta\gamma\langle i[B, C]\rangle]^{1/2}.$$

We now consider the left side of (2.2) with $\gamma \equiv \lambda\beta$ and $0 < \beta \ll 1$. From

(2.6) $\quad <A>' = <\psi|(I - i\beta DB + \gamma DC)A(I + i\beta DB + \gamma DC)|\psi> (\cos \theta)^2$

$$= <A> + \beta(<i[A, B]> + \lambda<\{DC, DA\}>) + O(\beta^2),$$

it follows that,

(2.7) $\quad |<A>' - <A>| = |\beta| \, |<i[A, B]> + \lambda<\{DC, DA\}>| + O(\beta^2)$,

where $O(\beta^n)$ will denote terms of order $\beta^n$ or higher. Similarly, we find,

(2.8) $\quad \Delta A' + \Delta A = 2 \Delta A + O(\beta)$.

Substituting (2.7, 8) into (2.2) and using (2.5b) for $\tan \theta$, we obtain,

(2.9) $\quad |\beta|( \, |<i[A, B]> + \lambda<\{DC, DA\}>| \, / \, 2\Delta A ) + O(\beta^2)$

$$\leq |\beta| [\Delta B^2 + \lambda^2 \Delta C^2 - \lambda <i[B, C]>]^{1/2} .$$

As this must hold for arbitrary small $|\beta|$, we must have,

(2.10) $\quad |<i[A, B]> + \lambda<\{DC, DA\}>|$

$$\leq 2\Delta A [\Delta B^2 + \lambda^2 \Delta C^2 - \lambda <i[B, C]>]^{1/2} ,$$

or, upon squaring both non-negative sides and moving all terms to the same side,

(2.11) $\quad \lambda^2 [4\Delta A^2 \Delta C^2 - <\{DC, DA\}>^2]$

$$- \lambda [4\Delta A^2 <i[B, C]> + 2<i[A, B]><\{DC, DA\}>] + [4\Delta A^2 \Delta B^2 - <i[A, B]>^2] \geq 0 .$$

For *this* inequality to hold for all real $\lambda$ we must have the constant term and the coefficient of $\lambda^2$ positive (unless all three coefficients vanish) and a non-positive discriminant, i.e.,

(2.12) $\quad [4\Delta A^2 <i[B, C]> + 2<i[A, B]><\{DC, DA\}>]^2$

$$- 4[4\Delta A^2 \Delta C^2 - <\{DC, DA\}>^2][4\Delta A^2 \Delta B^2 - <i[A, B]>^2] \leq 0 .$$

The surviving terms in the left side of (2.12) share a common factor of $16\Delta A^2$ and when that is factored out (permitted by the positivity of the coefficients of $\lambda^2$ and $\lambda^0$ in (2.11)) the resulting inequality can be written as,

(1.5) $\quad 4 \Delta A^2 \Delta B^2 \Delta C^2 \geq <i[A, B]>^2 \Delta C^2 + <i[B, C]>^2 \Delta A^2$

$$+ <\{DC, DA\}>^2 \Delta B^2 + <i[A, B]><i[B, C]><\{DC, DA\}>.$$

This result deserves a broader study than we will accord it here, focused as we are on the special case that yields (1.3). Accordingly, instead of moving directly to that special case, we pause, briefly, to comment on the relationship of (1.5) to some earlier uncertainty relations involving more than two observables.

Robertson (1934) considered uncertainty relations for arbitrary numbers of observables shortly after his classic work on two-observable uncertainty relations. All of his results were based on the non-negative character of the determinants of Hermitian matrices yielding positive definite quadratic forms. The matrices involved were the symmetric and anti-symetric parts of the matrices with mn elements of the form, $<DA_m\, DA_n>$, where the $A_m$ were the observables in question. However, for an odd number of observables the determinant of the anti-symetric matrix is identically zero. This and Robertson's interest in a connection he drew with related, classical, even dimensional phase space analyses resulted in his examining only the cases involving even numbers of observables.

Somewhat later J. L. Synge (1971) studied three-observable uncertainty relations employing methods related but not identical to those of Robertson. Synge obtains inequalities, which, with a little additional algebra, can be cast in the form,

(2.13a)  $\quad \Sigma[\Delta A^2(4\Delta B^2 \Delta C^2 - <i[B, C]>^2 - <\{DB, DC\}>^2)] \geq (3/4)|F| + (1/4)F$,

where the symbol, $\Sigma$, indicates a sum over the cyclic permutations of A, B and C in the expression following $\Sigma$ within the square brackets, and where,

(2.13b)  $\quad F = 8\Delta A^2 \Delta B^2 \Delta C^2 - <\{DA, DB\}><\{DB, DC\}><\{DC, DA\}>$

$\qquad\qquad + \Sigma[<i[A, B]><\{DB, DC\}><i[C, A]>]$.

The left side of (2.13a) involves a sum over quantities required to be non-negative by Robertson-Schroedinger. So if F is non-zero, (2.13a) strengthens at least one of the related Robertson-Schroedinger uncertainty relations.

More recently Trifonov (2002) has pursued uncertainty relations for arbitrary numbers of observables and states, but many of his results are already implied by the Robertson-Schroedinger uncertainty relations. In particular, Trifonov's single state, three observable uncertainty relation (equation (23) in Trifonov (2000)),

(2.14)  $\quad 2\Delta X^2[\Delta Y^2 + \Delta Z^2] \geq |<\{DX, DY\}><\{DX, DY\}>| + |<i[X, Y]><i[X, Z]>|$,

can readily be shown to follow from the two standard Robertson-Schroedinger uncertainty relations for (X, Y) and (X, Z).

While (2.13) is symmetric in A, B and C, (1.5) is not so. But a symmetric uncertainty relation can be obtained by simply adding together the cyclic permutations of (1.5). When this is done a result distinct from but similar to (2.13) is obtained, i.e.,

(2.15) $\quad \Sigma[\Delta A^2(4\Delta B^2\Delta C^2 - <i[B, C]>^2 - <\{DB, DC\}>^2)] \geq$

$\Sigma[\Delta A^2<i[B, C]>^2 + <i[A, B]><\{DB, DC\}><i[C, A]>] \equiv \| \Sigma[|\psi_A>\Delta A<i[B, C]>] \|^2 ,$

where, as Synge points out, for any self adjoint X defined on $|\psi>$ and with $\Delta X > 0$,

(2.16) $\qquad\qquad\qquad |\psi_X> \equiv DX|\psi>/\Delta X,$

is of unit norm and orthogonal to $|\psi>$. The squared vector norm on the rightmost side of (2.15), is non-zero unless $|\psi>$ satisfies the non-linear condition of being an eigenvector of the operator,

(2.17) $\qquad\qquad\qquad <i[B, C]>A + <i[C, A]>B + <i[A, B]>C.$

**3. (1.3) as a special case of (1.5):** The special case of (1.5) of interest to us here is that in which,

(3.1) $\qquad\qquad\qquad <[B, C]> = 0,$

i.e., B and C are effectively compatible. If (3.1) holds, then (1.5) becomes,

(3.2) $\qquad\qquad 4\Delta A^2\Delta B^2\Delta C^2 \geq <i[A, B]>^2\Delta C^2 + \Delta B^2<\{DC, DA\}>^2 .$

We note that if B and C are not merely effectively compatible, but identical, then (3.2) is just the Robertson-Schroedinger uncertainty relation, once the common factor of $\Delta B^2 = \Delta C^2$ is cancelled. From (2.11) this cancellation is permitted unless $<i[A, B]>^2 = <\{DC, DA\}>^2 = 0$, in which case it is pointless.

If we now express the anti-commutator term in (3.2) in terms of the CC between C and A, using (1.2), we have

(3.4) $\qquad\qquad 4\Delta A^2\Delta B^2\Delta C^2 \geq <i[A, B]>^2\Delta C^2 + 4\Delta A^2\Delta B^2\Delta C^2 K(A, C)^2 ,$

or, upon canceling the common factor, $\Delta C^2$,

(3.5a) $\qquad\qquad 4\Delta A^2\Delta B^2 \geq <i[A, B]>^2/(1 - K(A, C)^2) .$

Interchanging the roles of A and B in the preceding argument, we obtain,

(3.5b) $\qquad\qquad 4\Delta A^2\Delta B^2 \geq <i[A, B]>^2/(1 - K(D, B)^2),$

where,

(3.6) $\qquad\qquad\qquad <[A, D]> = 0.$

The promised result, (1.3/3.5), is established. But it has a puzzling aspect. In searching among the effective commutant of B we will often find some C for which, $K(A, C)^2 > K(A, B)^2$, and this will raise the lower bound on the (A, B) uncertainty product ratio for the state in question. But (1.3/3.5) also implies we will never find a C which will yield too large a squared correlation! Just what is it that keeps the squared correlations in check throughout the effective commutant? We will address that question in the next section by mounting a quite different derivation of (1.3/3.5). The alternative derivation is closer to the spirit of the standard Robertson-Schroedinger uncertainty relation than finding (1.3/3.5) as a special case of the three-observable uncertainty relation, (1.5). But while such a derivation was easily come by once (1.3/3.5) was in hand, it would not, otherwise, have occurred to the author.

**4. An alternative derivation of (1.3):** From (2.16) it follows that, for any self adjoint X and Y,

(4.1a)  $\qquad \langle i[X, Y]\rangle = -2\Delta X\, \Delta Y\, \text{Im}\langle \psi_X | \psi_Y \rangle$,

and

(4.1b)  $\qquad \langle \{DX, DY\}\rangle = 2\Delta X\, \Delta Y\, \text{Re}\langle \psi_X | \psi_Y \rangle$,

where 'Im' and 'Re' denote the imaginary and real parts, respectively, of the complex inner products following them.

Comparing (4.1b) with (1.2) we see that,

(4.1c)  $\qquad K(X, Y) = \text{Re}\langle \psi_X | \psi_Y \rangle$.

Applying (4.1a) and (4.1c) to the right hand sides of (1.3/3.5), we see that the inequalities, (1.3/3.5) are simply equivalent to,

(4.2a)  $\qquad (\text{Im}\langle \psi_A | \psi_B \rangle)^2 + (\text{Re}\langle \psi_A | \psi_C \rangle)^2 \leq 1$

and,

(4.2b)  $\qquad (\text{Im}\langle \psi_A | \psi_B \rangle)^2 + (\text{Re}\langle \psi_D | \psi_B \rangle)^2 \leq 1$.

These purely geometrical inequalities suggest a derivation of (1.3/3.5) having nothing explicit to do with UPRs or CCs and such a derivation follows. We focus on (4.2a).

Since the state vectors, $|\psi_A\rangle$, $|\psi_B\rangle$ and $|\psi_C\rangle$ are all of unit norm, it follows that if we write,

(4.3a)  $\qquad \langle \psi_A | \psi_B \rangle \equiv a + i\, b$,

(4.3b) $$\langle\psi_A|\psi_C\rangle \equiv x + i\, y,$$

with a, b, x and y real, then,

(4.4) $$a^2 + b^2 \leq 1 \quad \text{and} \quad x^2 + y^2 \leq 1.$$

Furthermore, from (3.1) and (4.1a), it follows that,

(4.5) $$\text{Im}\langle\psi_B|\psi_C\rangle = 0.$$

If we define $|\psi_{A;B}\rangle$ and $|\psi_{A;C}\rangle$ by,

(4.6a) $$|\psi_B\rangle \equiv |\psi_A\rangle\langle\psi_A|\psi_B\rangle + |\psi_{A;B}\rangle[1 - |\langle\psi_A|\psi_B\rangle|^2]^{1/2},$$

(4.6b) $$|\psi_C\rangle \equiv |\psi_A\rangle\langle\psi_A|\psi_C\rangle + |\psi_{A;C}\rangle[1 - |\langle\psi_A|\psi_C\rangle|^2]^{1/2}$$

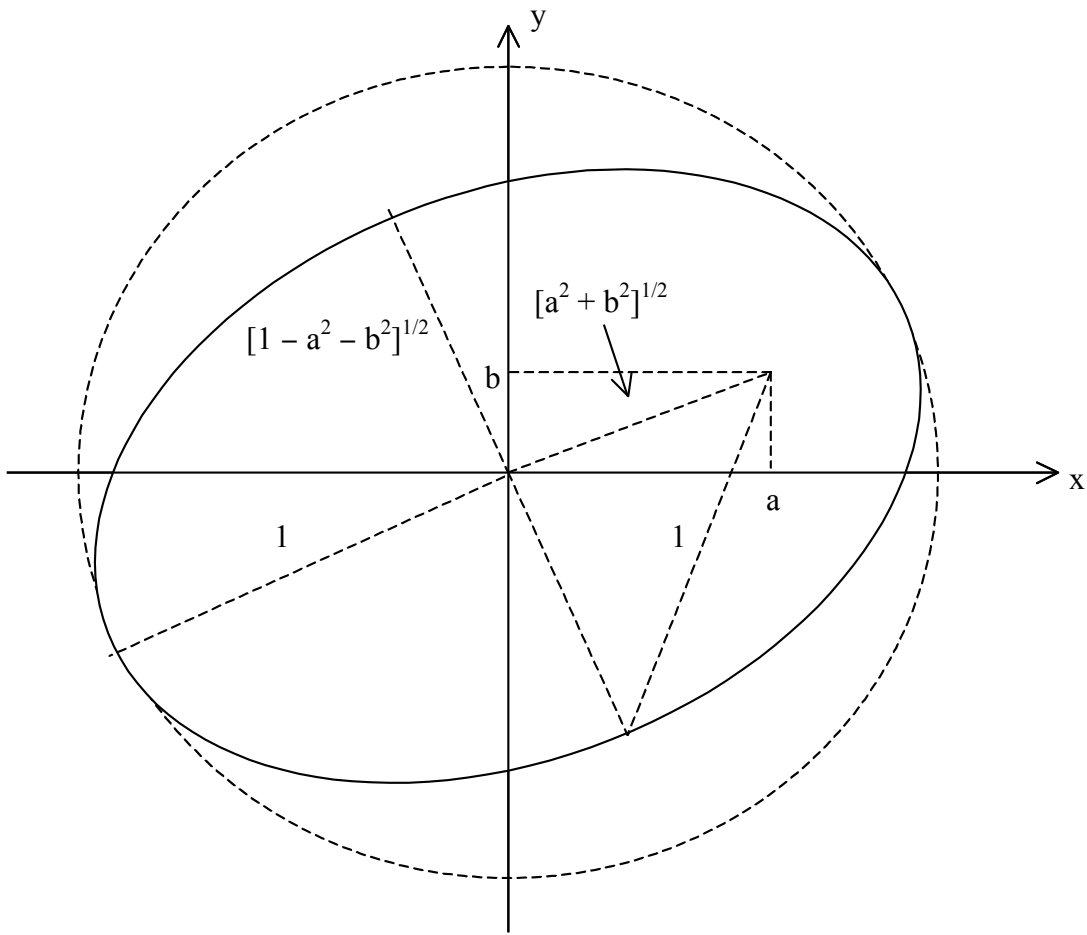

**Fig. 1:** Unit circle and confining ellipse in the complex (x, y) plane for $\langle\psi_A|\psi_C\rangle = x + i\, y$, where $\langle\psi_A|\psi_B\rangle = a + i\, b$ (with a, b > 0) and $\text{Im}\langle\psi_B|\psi_C\rangle = 0$.

then $|\psi_{A;\,B}\rangle$ and $|\psi_{A;\,C}\rangle$ are also of unit norm and orthogonal to both $|\psi\rangle$ and $|\psi_A\rangle$. Since we then have,

(4.7a)  $\quad\langle\psi_B|\psi_C\rangle = (a - i\,b)(x + i\,y) + \langle\psi_{A;\,B}|\psi_{A;\,C}\rangle(1 - a^2 - b^2)^{1/2}(1 - x^2 - y^2)^{1/2}$,

and

(4.7b)  $\quad 0 = \text{Im}\langle\psi_B|\psi_C\rangle = (a\,y - b\,x) + \text{Im}\langle\psi_{A;\,B}|\psi_{A;\,C}\rangle(1 - a^2 - b^2)^{1/2}(1 - x^2 - y^2)^{1/2}$,

it finally follows that,

(4.8a)  $\qquad\qquad (a\,y - b\,x)^2 \leq (1 - a^2 - b^2)(1 - x^2 - y^2)$,

or

(4.8b)  $\qquad\qquad (1 - a^2)\,x^2 - 2\,a\,b\,x\,y + (1 - b^2)\,y^2 \leq (1 - a^2 - b^2)$.

The *equality* in (4.8b) represents an ellipse in the x-y plane and the proper inequality represents the interior of the ellipse, (**Fig. 1**). Taking differentials of the left hand side of (4.8b) we obtain the differential equation for the ellipse,

(4.9)  $\qquad\qquad 2(1 - a^2)\,x\,dx - 2ab\,(y\,dx + x\,dy) + 2(1 - b^2)\,y\,dy = 0$.

The minimum and maximum values of x on the ellipse satisfy (4.9) when dx = 0, i.e., when,

(4.10)  $\qquad\qquad y = [ab/(1 - b^2)]\,x$.

Substituting (4.10) into (4.8b) yields, after a little algebra,

(4.11)  $\qquad\qquad b^2 + x^2 \leq 1$,

which, via (4.3), is just the desired (4.2a). Similar considerations yield (4.2b).

**5: Some examples:** (1st example) We begin with the example provided by the original Heisenberg uncertainty relation between position, x, and momentum, p, for a single quanton moving in one spatial dimension. When augmented by the Schroedinger strengthening and expressed in terms of the CC, $K(x, p)$ and the corresponding UPR, the uncertainty relation is,

(5.1)  $\qquad\qquad 2\,\Delta x\,\Delta p/\hbar \geq [1 - K(x, p)^2]^{-1/2}$.

If, now, f is an arbitrary real function of the observable, x, and g is an arbitrary real function of p, then our result, (1.3), asserts that,

(5.2a) $$2 \Delta x \, \Delta p / \hbar \geq [1 - K(f, p)^2]^{-1/2},$$

and

(5.2b) $$2 \Delta x \, \Delta p / \hbar \geq [1 - K(x, g)^2]^{-1/2}.$$

With no other degrees of freedom but the one spatial dimension, the class of functions of the form f and g exhaust the observables that commute with x and p, respectively.

To examine (5.2a), we express the position representation state function, $\psi(x)$ (suppressing the time dependence), as,

(5.3a) $$\psi(x) \equiv \psi_R(x) \, e^{i \phi(x)},$$

where

(5.3b) $$\psi_R(x) = \pm |\psi(x)|,$$

is chosen so as to maximize the continuity of the phase function, $\phi(x)$. We then find,

(5.4a) $$K(f, p)^2 = (\langle f \phi' \rangle - \langle f \rangle \langle \phi' \rangle)^2 / [\Delta f^2 (\Delta \phi'^2 + \langle (\psi_R'/\psi_R)^2 \rangle)],$$

where $\phi'$ and $\psi_R'$ indicate derivatives of $\phi$ and $\psi_R$, respectively. If $\phi$ is constant or varies with x no more than linearly, then all CCs of the form (5.4a) vanish. Then only (5.2b) might result in lifting the lower bound of the x-p UPR above unity. The minimum uncertainty wave packets are precisely those in which the CCs in (5.2b) also vanish, i.e., Gaussians with, at most, linearly dependent phases and similar Fourier transforms.

Variational considerations show that in the general case (5.4a) is maximized by the choice, $f = \phi'$, yielding,

(5.4b) $$K(f, p)^2{}_{\max} = K(\phi', p)^2 = \Delta \phi'^2 / [\Delta \phi'^2 + \langle (\psi_R'/\psi_R)^2 \rangle],$$

a result that is equivalent to Hall's (2001) identification of $\phi'(x)$ as the best estimator of p for a given value of x. As a consequence of (5.4b) we can assert,

(5.5) $$4 \Delta x^2 \Delta p^2 / \hbar^2 \geq [\Delta \phi'^2 + \langle (\psi_R'/\psi_R)^2 \rangle] / \langle (\psi_R'/\psi_R)^2 \rangle.$$

For the right hand side of (5.5) to be large enough for the equality to hold, the x-p UPR would have to be smaller than all the UPRs for observable pairs of the forms, (f, p) or ($\phi'$, g), where, as before, f and g are arbitrary functions of x and p, respectively.

Corresponding considerations apply for the CCs, $K(x, g)$, in (5.2b).

If the quanton in this example moved in three dimensional space and, perhaps, carried spin, **s**, so that p → $p_x$, and if the quanton was part of a composite system, then the functions, f and g, in the preceding discussion, would be arbitrary functions of the observable sets, (x, y, z, $p_y$, $p_z$, **s**, A) and (y, z, $p_x$, $p_y$, $p_z$, **s**, A), respectively, where A represents observables belonging to the environment of the quanton within the composite system.

($2^{nd}$ example) For our second example we consider a system of two quantons, each again moving in one dimension (alternatively, a single quanton moving in 2-space) for simplicity, and this time we specify the pure state with the unit norm, position representation state function,

(5.6) $\quad\quad\quad\quad \psi(x_1, x_2) = N \exp[- X^2/4a^2] \exp[- r^2/4b^2]$,

where, $X \equiv (x_1 + x_2)/2$ and $r \equiv (x_1 - x_2)$. Being a product of minimum uncertainty Gaussian wave packets in X and r centered on 0 we have,

(5.7a) $\quad\quad\quad <X> = <r> = <P> = <p> = <X\,r> = <X\,p> = <P\,r> = <P\,p>$

$$= <XP + PX> = <rp + pr> = 0$$

and

(5.7b), $\quad\quad\quad \Delta X^2 = a^2, \quad \Delta P^2 = \hbar^2/4a^2, \quad \Delta r^2 = b^2, \quad \Delta p^2 = \hbar^2/4b^2$,

where $P \equiv p_1 + p_2$ and $p \equiv (p_1 - p_2)/2$ are canonically conjugate to X and r, respectively. It follows that,

(5.8a) $\quad\quad \Delta x_{1,2}^2 = <x_{1,2}^2> = <(X \pm (r/2))^2> = \Delta X^2 + \Delta r^2/4 = a^2 + b^2/4$,

and

(5.8b) $\quad\quad \Delta p_{1,2}^2 = <p_{1,2}^2> = <((P/2) \pm p)^2> = (\Delta P^2/4) + \Delta p^2 = (\hbar^2/4)[(1/4a^2) + (1/b^2)]$,

yielding the squared UPR,

(5.8c) $\quad\quad\quad\quad 4\Delta x_1^2 \Delta p_1^2/\hbar^2 = [(2a/b) + (b/2a)]^2 / 4$.

If (b/2a) is either much larger or much smaller than unity, then the squared UPR is much larger than unity. But Schroedinger's modification of the Heisenberg uncertainty relation for $x_1$ and $p_1$ is of no help in indicating this large value since, as follows from (5.7a), the CC, $K(x_1, p_1)$, is zero.

However, our (1.3) allows us to use $K(x_1, x_2)$ in place of $K(x_1, p_1)$, and when we do we find,

(5.9a) $\quad K(x_1, x_2)^2 = <x_1x_2>^2/\Delta x_1^2 \Delta x_2^2 = [(2a/b) - (b/2a)]^2 / [(2a/b) + (b/2a)]^2$

Consequently,

(5.9b) $\quad [1 - K(x_1, x_2)^2]^{-1} = [(2a/b) + (b/2a)]^2 / 4,$

equaling the squared UPR (5.8c)! According to (1.3) no other CC that could have been used here has a square larger than $K(x_1, x_2)^2$ and no other UPR for which $K(x_1, x_2)^2$ could be used is smaller than (5.8c).

(3$^{rd}$ example) Here we examine our result (1.3) for the case of a bi-partite system in a general entangled state expressed via the bi-orthogonal decomposition,

(5.10a) $\quad |\Psi> \equiv \Sigma_n a_n |\phi_n>|\chi_n>,$

where all the $a_n > 0$,

(5.10b) $\quad <\phi_m|\phi_n> = <\chi_m|\chi_n> = \delta_{mn} \quad$ and $\quad \Sigma_n a_n^2 = 1.$

We will call the $|\phi_n>$, basis states for the *first* subsystem and the $|\chi_n>$, basis states for the *second* subsystem. These states may not span the respective factor spaces for the subsystems and we will have use for the factor space projectors,

(5.10c) $\quad \Pi_1 \equiv I_1 - \Sigma_m |\phi_m><\phi_m|, \quad \Pi_2 \equiv I_2 - \Sigma_m |\chi_m><\chi_m|,$

where the $I_1$ and $I_2$ are the factor space identity operators.

We consider a UPR for two observables, A and B, belonging to the first subsystem and we ask for the maximum squared CC that can exist between A and any observable, C, belonging to the second subsystem. Such a C automatically commutes with B and so, by (1.3), can be used to infer a lower limit for the UPR. Maximizing such restricted squared CCs is not guaranteed to produce the highest squared CC useable in (1.3) since some observable, E, belonging to the first subsystem or belonging to neither subsystem, but which commutes with B, may yield the highest squared CC. Nevertheless, it is interesting to examine this restricted class.

We have,

(5.11) $\quad K(A, C)^2 = [<AC> - <A><C>]^2/\Delta A^2 \Delta C^2.$

and for the state, (5.10), we have,

(5.12a) $\quad <C> = \Sigma_m a_m^2 C_{mm}, \quad <C^2> = \Sigma_{m,n} a_m^2 C_{mn}C_{nm} + \Sigma_m a_m^2 (C\Pi_2 C)_{mm},$

and,

(5.12b) $$\langle AC\rangle = \Sigma_{m,n}\, a_m\, A_{mn} C_{mn}\, a_n\,,$$

where,

(5.12c) $$A_{mn} = \langle\phi_m|A|\phi_n\rangle\,, \qquad C_{mn} = \langle\chi_m|C|\chi_n\rangle.$$

Defining,

(5.13) $$N \equiv \langle AC\rangle - \langle A\rangle\langle C\rangle,$$

We have, for (5.11),

(5.14) $$\delta K(A, C)^2 = 2N\delta N/\Delta A^2 \Delta C^2 - N^2 \delta\Delta C^2/\Delta A^2 \Delta C^4$$

$$= (N/\Delta A^2 \Delta C^2)[2\delta N - (N/\Delta C^2)\delta\Delta C^2].$$

The squared CC, (5.11), is maximized when $2\delta N - (N/\Delta C^2)\delta\Delta C^2 = 0$, i.e.,

(5.15a) $$2[\Sigma_{m,n}\, a_m\, A_{mn}\delta C_{mn}\, a_n - \langle A\rangle \Sigma_m\, a_m^2\, \delta C_{mm}]$$

$$= (N/\Delta C^2)[\Sigma_{m,n}\, a_m^2\, (C_{mn}\delta C_{nm} + \delta C_{mn} C_{nm}) - 2\langle C\rangle \Sigma_m\, a_m^2\, \delta C_{mm} + \Sigma_m\, a_m^2 \delta(C\Pi_2 C)_{mm}]$$

or

(5.15b) $$2[\Sigma_{m,n}\, a_m(A_{mn} - \langle A\rangle \delta_{mn})\, a_n\, \delta C_{mn}]$$

$$= (N/\Delta C^2)[\Sigma_{m,n}\, (a_m^2 + a_n^2)(C_{nm} - \langle C\rangle\delta_{mn})\delta C_{mn} + \Sigma_m\, a_m^2(\delta C\Pi_2 C + C\Pi_2 \delta C)_{mm}\,,$$

for independent and arbitrary infinitesimal variations $\delta C_{mn} = \delta C_{nm}{}^*$ and ${}_m\delta C\Pi_2 = \Pi_2\delta C_m{}^\dagger$. Consequently, the values of the $C_{mn}$ which maximize the squared CC are given by,

(5.16) $$C_{mn} - \langle C\rangle\delta_{mn} = [2\, a_m a_n/(a_m^2 + a_n^2)](A_{mn}{}^* - \langle A\rangle\delta_{mn})$$

$$= \lambda\{[2\, a_m a_n/(a_m^2 + a_n^2)]A_{mn}{}^* - \langle A\rangle\delta_{mn}\}\,,$$

where $\lambda \equiv \Delta C^2/N$, while the $\Pi_2 C_m = {}_m C\Pi_2{}^\dagger = 0$.

Substitution of these values yields,

(5.17a) $$N = \langle(A - \langle A\rangle)(C - \langle C\rangle)\rangle = \Sigma_{m,n}\, a_m(A_{mn} - \langle A\rangle\delta_{mn})(C_{mn} - \langle C\rangle\delta_{mn})a_n$$

$$= \lambda\,\{\Sigma_{m,n}\,[2\, a_m^2 a_n^2/(a_m^2 + a_n^2)]|A_{mn}|^2 - \langle A\rangle^2\},$$

and

(5.17b) $$\Delta C^2 = \langle (C - \langle C \rangle)^2 \rangle = \Sigma_{m,n}\, a_m^2\, |C_{mn} - \langle C \rangle \delta_{mn}|^2$$

$$= \lambda^2 \{ \Sigma_{m,n}\, a_m^2 [4\, a_m^2 a_n^2/(a_m^2 + a_n^2)^2]|A_{mn}|^2 - \langle A \rangle^2 \}$$

$$= \lambda^2 \{ \Sigma_{m,n}\, [2\, a_m^2 a_n^2/(a_m^2 + a_n^2)]|A_{mn}|^2 - \langle A \rangle^2 \} \, .$$

Therefore,

(5.18) $$K(A, C)^2_{max} = N^2/\Delta A^2 \Delta C^2 = N / \lambda \Delta A^2$$

$$= \{ \Sigma_{m,n}\, [2\, a_m^2 a_n^2/(a_m^2 + a_n^2)]|A_{mn}|^2 - \langle A \rangle^2 \}/\Delta A^2.$$

But,

(5.19a) $$\Delta A^2 = \langle A^2 \rangle - \langle A \rangle^2 = \Sigma_{m,n}\, a_m^2 |A_{mn}|^2 - \langle A \rangle^2 + \Sigma_m\, a_m^2 (A\Pi_1 A)_{mm}$$

$$= \Sigma_{m,n}\, (1/2)(a_m^2 + a_n^2)|A_{mn}|^2 - \langle A \rangle^2 + \Sigma_m\, a_m^2 (A\Pi_1 A)_{mm},$$

and,

(5.19b) $$[2\, a_m^2 a_n^2/(a_m^2 + a_n^2)] = (1/2)(a_m^2 + a_n^2) - (a_m^2 - a_n^2)^2/2(a_m^2 + a_n^2).$$

Substituting (5.19a, b) into (5.18) we can rewrite (5.18) as,

(5.20) $$K(A, C)^2_{max}$$

$$= 1 - \Delta A^{-2} \{ \Sigma_{m,n}[(a_m^2 - a_n^2)^2/2(a_m^2 + a_n^2)]|A_{mn}|^2 + \Sigma_m\, a_m^2 (A\Pi_1 A)_{mm} \} \, ,$$

the maximum squared CC between A, belonging to the *first* system, and *any* observable belonging to the *second* system. We note that if the positive coefficients, $a_n$, are all equal and if the $(A\Pi_1 A)_{mm}$ are all zero then the maximum squared CC equals unity and the UPR for A and B in subsystem 1 is forced, by (1.3), to be infinite. But the specified conditions are just sufficient to guarantee that the expectation value of the commutator between A and B is zero and the UPR *is* infinite.

**References:**

Fleming, G. N. (2001) "Uses of a Quantum Master Inequality", available at philsci-archive.pitt.edu/archive/00000646 and at arxiv.org/physics/0106077.

Hall, M. J. W. (2001) "Exact uncertainty relations", available at quant-ph/0107149.

Robertson, H. P. (1929) "The uncertainty principle", *Physical Review* **34**, pp. 163-4.
-------------------- (1930) "A general formulation of the uncertainty principle and its classical interpretation", *Physical Review* **35**, p. 667A.
-------------------- (1934) "An Indeterminacy Relation for Several Observables and its Classical Interpretation", *Physical Review* **46**, pp. 794-801.

Schroedinger, E. (1930) "Zum Heisenbergschen Unschärfeprinzip", *Berliner Berichte 1930*, pp. 296-303. English translation available at quant-ph/9903100.

Synge, J. L. (1971) "Geometrical approach to the Heisenberg uncertainty relations and its generalizations", *Proc. R. Society, London,* A. **325**, pp. 151-156.

Trifonov, D. A. (2000) "State Extended Uncertainty Relations" J. Phys. A **33**, L299-L304. e-print at quant-ph/0005086
-------------------- (2002) "Generalizations of Heisenberg Uncertainty Relations", e-print at quant-ph/0112028